\documentclass[12pt]{article}
\usepackage{epsfig}
\usepackage{amssymb}

\def\beq{\begin{equation}}
\def\eeq#1{\label{#1}\end{equation}}
\def\eeqn{\end{equation}}

\def\beqa{\begin{eqnarray}}
\def\eeqa#1{\label{#1}\end{eqnarray}}
\def\eeqan{\end{eqnarray}}

\let\bar=\overbar

\def\Dslash{\not{\hbox{\kern-4pt $D$}}}
\def\dslash{\not{\hbox{\kern-2pt $\del$}}}

\def\msb{{\bar{\ssstyle M \kern -1pt S}}}

\usepackage{fancyhdr,graphicx}
\def\Title#1{\begin{center} {\Large {\bf #1} } \end{center}}

\begin{document}

\Title{Cooling and Heating Solid Quark Stars}

\bigskip\bigskip


\begin{raggedright}

{\it Meng Yu}\\
School of Physics and State Key Laboratory of Nuclear Physics and
Technology\\
Peking University\\
Beijing 100871\\
P. R. China\\
{\tt Email: vela.yumeng@gmail.com}
\bigskip\bigskip
\end{raggedright}

\section{Introduction}
The study of cold quark matter has been an interesting topic of research in recent years. In an astrophysical
context,  quark matter may be used to explain the measureable properties of pulsar-like compact stars.

It has recently been proposed that realistic quark matter in compact stars could exist in a solid state
~\cite{xu03,Horvath05}, either as a super-solid or as a normal solid~\cite{xu09}. The basic conjecture of normal
solid quark matter is that de-confined quarks tend to form quark-clusters when the temperature and density are
relatively low. Below some critical temperature, these clusters could be in periodic lattices immersed in a
degenerate, extremely relativistic, electron gas. Note that even though quark matter is usually described as
weakly coupled, the interaction between quarks and gluons in a quark-gluon plasma is still very
strong~\cite{Shuryak}. It is this strong coupling that could cause the quarks to cluster and form a solid-like
material.

In this paper, we argue that comtemporary observations of thermal X-ray emitting pulsars are consistent with the
assumption that these sources are in fact Solid Quark Stars (SQSs).
In the successive Sect., the thermal X-ray observations on isolated pulsars are exhibited according to the
collation in this work.
In Sect. 3, the SQS pulsar model is explored to interpret these observations.
Conclusions and corresponding discussions are presented in Sect. 4

\section{The observations}
\begin{figure}
\begin{center}
\epsfig{file=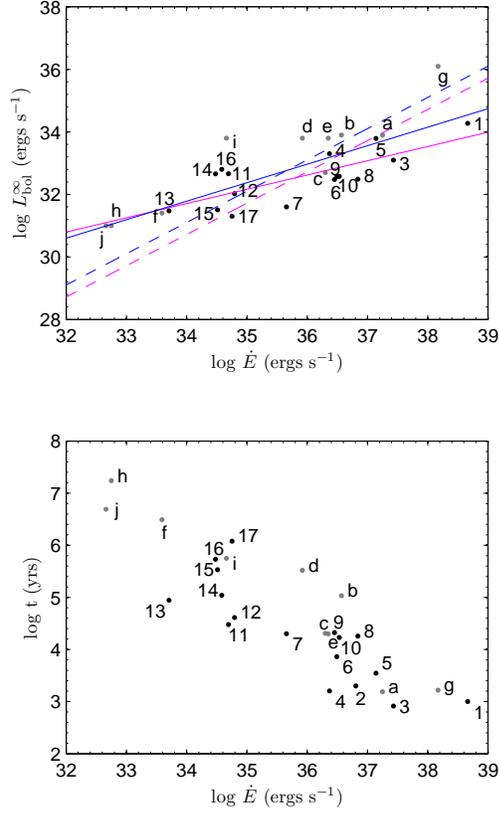,height=4.8in} \caption{Functions $L_{\rm bol}^{\infty}(\dot{E})$ ({\it top} panel) and
$t(\dot{E})$ ({\it bottom} panel) of active pulsar candidates. In the {\it top} panel, the fits are carried out
both for Group A (including top 17 pulsars listed in Table 1 in \cite{yx09}; they are marked by dark points and
their numbers) and Group B (including the members in Group A and the other 10 sources whose upper limits on the
bolometric luminosities could be defined observationally; the 10 sources are marked by grey points and letters).
Red lines are the fitted results for Group A, while blue lines are those for Group B. In both groups, solid
lines provide the best fits to the data, while the dashed lines give the fits by freezing $p1$ at 1. We note
that the 10 sources with upper limits on their bolometric luminosities defined are taken from ~\cite{BA02}, and
they are a. B1509-58, b. B1951+32, c. B1046-58, d. B1259-63, e. B1800-21, f. B1929+10, g. B0540-69, h. B0950+08,
i. B0355+54, j. B0823+26.}\label{fig:edot}
\end{center}
\end{figure}
\begin{table}
{\scriptsize
\begin{center}
\begin{tabular}{ccccc}
\hline \\
Group & $p1$$^a$ & $p2$ & Corr. Coef.$^b$ & $\chi_r^2$ (d.o.f)$^c$ \\
\hline \\
A. & $0.4561\pm0.2315$ & $16.20\pm8.30$ & -- & 0.3277(14) \\
 & (1) & $-3.280^{+0.495}_{-0.494}$ & 0.7487 & 0.8607(15) \\
B. & $0.5918^{+0.2045}_{-0.2046}$ & $11.66^{+7.3100}_{-7.2990}$ &
-- & 0.6081(24) \\
 & (1) & $-2.896\pm0.403$ & 0.7730 & 0.9964(25) \\
\hline \\
\end{tabular}
\end{center}
\begin{flushleft}
$^a$ Values in the parenthesis are frozen during the fits. The errors are in 95\% confidence level.
\newline
$^b$ Correlation coefficient of the data between bolometric luminosity and $\dot{E}$.\newline
$^c$ Reduced $\chi^2$, or $\chi^2$ per degree of freedom (d.o.f).
\caption{Fitting parameters for $L_{\rm bol}^{\infty}$---$\dot{E}$ data, which has been shown in Fig.
\ref{fig:edot} ({\it top} panel).}\label{tab:fit}
\end{flushleft}}
\end{table}

We collate the thermal X-ray observations on 29 isolated pulsars (see Table 1 in~\cite{yx09}). The black body
parameters summarized comprise of pulsar surface temperature components $T_{\rm
s,1/2}^{\infty}$\footnote{``$\infty$''---values measured at distant.}, emission size components
$R_{1/2}^{\infty}$, ages and bolometric X-ray luminosities $L_{\rm bol}^{\infty}$.
Spin would be a significant source of energy for pulsars. Potential relation between pulsar X-ray
bolometric luminosity $L_{\rm bol}^{\infty}$ and its spin energy loss rate $\dot{E}$ could exist. Such a study
was done and the results are exhibited in Fig. \ref{fig:edot} and Table \ref{tab:fit}.
Using the function of
\begin{equation}
{\rm log}L_{\rm bol}^{\infty}=p1\cdot {\rm log}\dot{E}+p2, \label{eq:fit}
\end{equation}
$L_{\rm bol}^{\infty}-\dot{E}$ data were fitted and the results show that the best fit demonstrates a possible 1/2-law between
these two quantities, or
\begin{equation}
L_{\rm bol}^{\infty}(\dot{E})=C\dot{E}^{1/2}, \label{eq:onesecond}
\end{equation}
where the coefficient $C=10^{p2}$ is in the unit of ${\rm erg^{1/2} s^{-1}}$.
However, when the parameter $p1$ is set to 1, a linear-law of
\begin{equation}
L_{\rm bol}^{\infty}(\dot{E})=\eta\cdot\dot{E}, \label{eq:linear}
\end{equation}
is presented, and this coefficient $\eta$ is $10^{-3}$ approximately.

\section{The SQS pulsar model}
\subsection{X-ray cooling and accreting pulsars}
The state of a pulsar would be manifested in multiple wave bands and by different photon components. Top 17
pulsars have significant observable X-ray non-thermal components, which could originate from energetic
magnetospheres. Such pulsars are considered to be undergoing cooling processes as proposing by SQSs.
The rest sources are dominated by thermal X-ray components observationally, and show little magnetospheric
activities. These pulsars may be accreting to the surrounding interstellar medium to generate visible X-ray
luminosities, if they are in fact SQSs.
\subsection{Cooling of SQSs}
\begin{figure}[htb]
\begin{center}
\epsfig{file=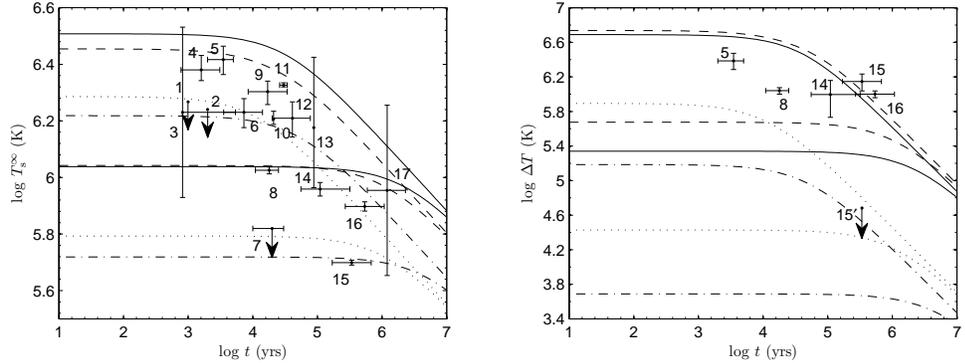,height=2.0in} \caption{{\it Left} panel: Cooling curves for SQSs, if the 1/2-law between the
bolometric luminosity and the spin energy loss rate holds. {\it Right} panel: Corresponding temperature
differences between the hot and warm components of SQSs, or $\Delta T=T^{\infty}_{\rm p}-T^{\infty}_{\rm s}$.
The parameters in both panels: $M=0.1{\rm M_\odot}$, $C=10^{16}$ (solid lines); $M=1.0{\rm M_\odot}$,
$C=10^{16}$ (dashed lines); $M=1.0{\rm M_\odot}$, $C=10^{15}$ (dotted lines); $M=0.01{\rm M_\odot}$, $C=10^{15}$
(dash-dot lines) ($C$ is in the unit of erg$^{1/2}$ s$^{-1/2}$). For two curves with same $M$ and $C$, the upper
one corresponds to an initial spin of 10 ms, while that of the lower one is 100 ms. Noting that the errors on
the surface temperatures of PSRs J0205+6449 (No. 3) and J2043+2740 (No. 17) are not provided by the authors of
the references, we then conservatively adopt them as deviating from the central values by a factor of
2.}\label{fig:ccos}
\end{center}
\end{figure}
\begin{figure}[htb]
\begin{center}
\epsfig{file=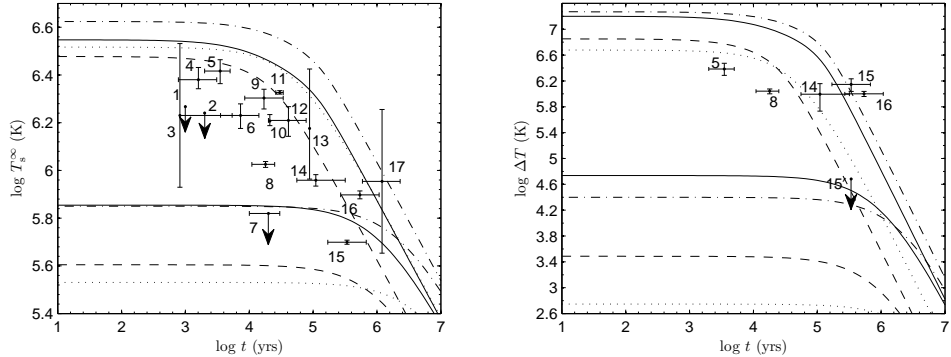,height=2.0in} \caption{{\it Left} panel: Cooling curves for SQSs, if the linear-law holds.
{\it Right} panel: Temperature differences for this case. The parameters: $M=1.0{\rm M_\odot}$, $\eta=0.01$
(solid lines); $M=1.0{\rm M_\odot}$, $\eta=0.001$ (dashed lines); $M=0.1{\rm M_\odot}$, $\eta=0.1$ (dash-dot
lines); $M=0.01{\rm M_\odot}$, $\eta=0.1$ (dotted lines). As in Fig. \ref{fig:ccos}, for two curves with same
$M$ and $\eta$, the upper one corresponds to an initial spin of 10 ms, while the lower one 100
ms.}\label{fig:cclr}
\end{center}
\end{figure}
Cooling of a quark star, from the initial birth, can be quite complicated to model.
Two facts induce this result. Firstly, recent experiments on collisions between heavy relativistic ions (e.g.
RHIC) manifested that the potential quark-gluon plasma in the fire ball would be strongly coupled even in the
temperatures of a few factors of the critical temperature $\sim$170 MeV~\cite{Shuryak}. However, and secondly,
the initial temperature at the birth of a pulsar could only be a few factors of $\sim$ 10 MeV. Hence, if the
natal pulsar is a quark star, the stellar interior quark-gluon plasma would be coupled strongly, so that current
physics could not model either in an analytic or a
numerical way.

Nevertheless, a phenomenological study on this problem could be carried out, and Fig. 1 in~\cite{yx09}
illustrates proposed cooling processes for a quark star. It could be quite uncertain that at which temperatures
the phase transitions would occur, and how long the star would stay in Stages 1 and 2.
However, in consequence of the strong coupling in quark-gluon plasma, Stages 1 and 2 would not be long, and the
observed pulsars (if they are quark stars) would be in the solid phase (Stage 3), because of the measured low
temperatures.
The calculations below concentrate on the late Stage 3, where the surface temperature is in the range of the
observed $10^7$---$10^5$ K. The loss of the thermal energy of a SQS in the late 3rd stage could dominantly induced by the emission of
photons rather than neutrinos, as a result of the low neutrino emissivities in such low temperatures.

In general, the cooling process would be described by the equation of
\begin{equation}
-C_v\frac{{\rm d}T_{\rm s}}{{\rm d}t}+L_{\rm SH}= L_{\rm bol}, \label{eq:coolinggeneral}
\end{equation}
with assuming that the stellar volume is invariant and the internal energy of the star is only the function of
the stellar temperature $T_{\rm s}$. $C_v$ is the stellar heat capacity. $L_{\rm SH}$ is the luminosity of the
probable heating process, and $L_{\rm bol}$ is the photon luminosity of the star.
The stellar heat capacity was found to be tiny, so that the internal energy would be far from sustaining
long-term X-ray emission. We note here that for a $1.4{\rm M_{\odot}}$-SQS, it would cool down to 1 K from
$10^{11}$ K in $\sim$0.2 s. The detailed exhibition of the calculation of this stellar residual internal energy
as well as heat capacity can be found in Appendix A.

Heating mechanism would take important effect during the cooling of SQSs. In such a way, the thermal evolution
of a SQS could not be a self-sustaining cooling process, but an external-supplying heating process.
The bombardment by backflowing particles at the poles could be the prime heating mechanism for SQSs, which would
intrinsically be spin-powered.
Combining the negligible residual internal energy of a SQS and the observed $L_{\rm bol}^{\infty}$---$\dot{E}$
relations. The heating luminosity $L_{\rm SH}$ could also follow that 1/2-law or linear-law. The equation
(\ref{eq:coolinggeneral}) would hence yields
\begin{equation}
C\dot{E}^{1/2}=4\pi R^2\sigma T_{\rm s}^{4}+4\pi r_{\rm p}^2\sigma T_{\rm p}^4, \label{eq:coolingactiveos}
\end{equation}
or
\begin{equation}
\eta\dot{E}=4\pi R^2\sigma T_{\rm s}^{4}+4\pi r_{\rm p}^2\sigma T_{\rm p}^4. \label{eq:coolingactivelr}
\end{equation}
The two terms on the right side of the equal sign are the stellar primary black body emission component and that
for a probable hot pole component respectively. The relation between $T_{\rm p}$ and $T_{\rm s}$ could be found
by
\begin{equation}
H\sim \frac{L_{\rm SH}-4\pi r_{\rm p}^{2}\sigma T_{\rm p}^{4}}{\pi r_{\rm p}^{2}}\sim \kappa_{\rm e}\frac{T_{\rm
p}-T_{\rm s}}{R}, \label{eq:H}
\end{equation}
where $H$ is the heat current flowing from the poles to the bulk of the star. The stellar heat conductivity
$\kappa_{\rm e}$ mainly contributed by free electrons refers to the corresponding results for solid
metal~\cite{FI81}.
Figs. \ref{fig:ccos} and \ref{fig:cclr} exhibit the computed cooling curves for the above two cases
respectively. In these calculations, SQSs were assumed to slow down as orthogonal rotators that lose their spin
energy by magnetic dipole radiation. The average densities of the SQSs were adopted to be 3 times nuclear
saturation density. Note that the results will not vary significantly when the density values are in the range of 3--5 nuclear saturation density, which is thought to be representative values for the solid quark matter.
The spin-powered thermal emission of SQSs enable an estimate on pulsar moments of inertia~\cite{yx09}.
\subsection{Accreting SQSs}
Dim X-ray isolated neutron stars (XDINs) and compact central objects (CCOs) (Sources No. 18-29 and 4) are
thought to be inactive pulsar candidates with inert magnetospheric manifestations. These sources may have
excessive X-ray luminosities to their spin energy loss rates.
Their energy origin has been keeping disputable, but accretion origin would be a candidate. The relative high
magnetic fields of these sources may drive their surrounding shells of matter (in large radii, usually larger
than the corotation radii) to rotate together with them. Under this {\it propeller} regime, nearly all of the
matter would be expelled outward, but a small fraction may diffuse starward and fall onto the stellar surfaces
eventually. The release of the gravitational energy as well as the latent heat inducing by the burning from the
baryonic matter phase to the three-flavor quark matter phase could sustain long-term soft X-ray radiations of
such pulsars.
In this scenario, the stellar luminosity would then be
\begin{equation}
L_{\rm bol}=\frac{GM\cdot \eta_{\rm acc}\cdot\dot{M}}{R}+\Delta\varepsilon\frac{\eta_{\rm
acc}\cdot\dot{M}}{m_{\rm p}}. \label{eq:coolingdead}
\end{equation}
For the calculated accreting parameters for SQSs in this {\it propeller} regime and the resulting luminosities,
please see \cite{yx09}.
Note here that for No. 4 RX J0822-4300, a CCO, we treated it both as a cooling SQS and a accreting SQS, as the
result of the disputable structure around it~\cite{ZTP99,HB06}.

\section{Conclusions}
We collate the thermal observations of 29 X-ray isolated pulsars and, in the SQS regime, for the
magnetospherically active pulsar candidates, establish their cooling processes (Figs. \ref{fig:ccos} and
\ref{fig:cclr}), while for the magnetospherically inactive or dead pulsar candidates, interpret the X-ray
luminosities under the accretion scenario.
SQSs, because of the possibility of being low-mass, could provide an approach to understand the small black body
emission sizes. We note that for SQSs with mass of $0.01{\rm M_\odot}$, ${\rm 0.1{\rm M_\odot}}$ and $1{\rm
M_\odot}$, their radii are $\sim$1.8, $\sim$3.8 and $\sim$8.3 km respectively.
On the other hand, a linkage between pulsar rotational kinetic energy loss rates and bolometric X-ray
luminosities is explored by SQSs.

We hence conclude that the phenomenological SQS pulsar model could not be ruled out by the thermal observations
on X-ray isolated pulsars, though a full depiction on the formation and the thermal evolution for a quark star
in all stages could hardly be given nowadays, as the lack of the physics in some extreme conditions (as has been
discussed in Sect. 3.2).
SQSs have significant distinguishable interiors with neutron stars, but the structures of the magnetosphere
between these two pulsar models could be similar. Therefore, SQSs and neutron stars would have similar heating
mechanisms, such as the bombardment by backflowing particles and the accretion to surrounding medium. However, a
full comparison between SQSs and neutron stars including their cooling processes as well as heating mechanisms
has beyond the scope of this paper, and should be interesting and necessary in the future study.

\appendix

\section{Residual internal energy of a SQS---fast cooling?}
Provided that the volume of a SQS is a constant, then the stellar residual thermal energy $U_{\rm SQS}$ would
only be the function of the stellar temperature $T_{\rm s}$,
\begin{equation}
U_{\rm SQS}(T_{\rm s})=\int C_{v}{\rm d}T_{\rm s}, \label{eq:inner}
\end{equation}
where $T_{\rm s}$ is the value in the star's local reference frame,
and $C_{v}$ is the heat capacity of the star.
Using Debye elastic medium theory, the characteristic of the heat capacity of solid quark matter would be
evaluated by Debye temperature $\theta_{D}$, or
\begin{equation}
\theta_{D}=\hbar \omega_D/k_{\rm B},
\end{equation}
where $\hbar$ is reduced Planck constant, and $k_{\rm B}$ is Boltzmann constant.
$\omega_D$ is Debye cut-off frequency (i.e. the maximum frequency of the wave that could propagate in a medium),
which equals Debye wave number $k_{D}=(6\pi^2n_{\rm c})^{1/3}$ times the average sound speed in the medium, i.e.
$\omega_D=k_{D}\bar{c}_s$, where $n_{\rm c}$ is the number density of classical particles, or quark clusters for
solid quark matter.
For a SQS, the average sound speed could be the light speed approximately. A linear equation of state, extended
to be used for a quark star, indicates that the pressure $p\sim \rho c^2$, where $\rho$ is the mass density of a
quark star and $c$ is the speed of light. So an estimate would be $\bar{c}_s=\sqrt{{\rm d}p/{\rm d}\rho}\sim
\sqrt{p/\rho}\sim \sqrt{\rho c^2/\rho}=c$.
$n_{\rm c}=3\epsilon n_0/A$, where $\epsilon$ denotes the baryon number density of solid quark matter in the
unit of $n_0$. $n_0=0.17$ fm$^{-3}$ is the nuclear saturation baryon number density. We consider $\epsilon=3-5$
could be the representative values for a SQS. $A$ is the number of valence quarks in a quark cluster. We may
expect that $A$ could be in the order of 10, since $A=18$ if quark-$\alpha$-like clusters are
formed~\cite{xu03}, and $A$ could even be conjectured to be in the order of $\geq 10^2$.
Debye temperature $\theta_D$ of a SQS could then be of order $\sim 10^{12}$ K, even much higher than the
temperature when a quark star is born. Hence, the heat capacity in the low-temperature limit or the
temperature-cube law, would be applicable for the SQS stage (Stage 3), i.e.
\begin{equation}
c_v = \frac{12\pi^4}{5}k_{\rm B}(\frac{T_{\rm s}}{\theta_D})^{3}
\label{eq:cv},
\end{equation}
where $c_v$ is the heat capacity per classical particle (or quark cluster when refers to the solid quark
matter). Thus $C_v=N\cdot c_v$, where $N$ is the total number of clusters in a star.

By referring to equation (\ref{eq:inner}), the residual internal energy for a 1.4${\rm M_\odot}$-SQS would thus
be $U_{\rm SQS}\sim10^{4}T_{\rm s}^4$ erg, while the corresponding value for a typical neutron star in such mass
is $U_{\rm NS}\sim10^{30}T_{\rm s}^2$ erg ~\cite{yak04}. Therefore, the residual internal energy of a SQS will
always be much less than that of a normal neutron star; even when the temperature is as high as $\sim10^{11}$ K,
the residual thermal energy of a SQS is only one ten thousandth of that of a neutron star. So if the cooling
process of a 1.4${\rm M_\odot}$-SQS is only powered by its residual internal energy, i.e.
\begin{equation}
-C_v\frac{{\rm d}T_{\rm s}}{{\rm d}t}=4\pi R^2\sigma T_{\rm s}^4+\frac{4}{3}\pi R^310^{24}T_{\rm s,9}^6
\label{eq:timesl}
\end{equation}
($R$ is the local stellar radius, $\sigma$ is Stefan-Boltzmann constant), then its cooling timescale from
$10^{11}$ K to an extremely low final temperature 1 K is only $\sim0.2$ s.
Noting that as a result of the concerning of the high initial temperature here, neutrino luminosity for the SQS
is also considered as the second term on the right side of the equal sign in equation (\ref{eq:timesl}).

\bigskip
The authors are grateful to Prof. Fredrick Jenet for his valuable comments and advices. We wish to thank Mr.
Weiwei Zhu for his detailed introduction to the observational situation about the sources PSR J1811-1925 and PSR
B1916+14. The authors also thank the colleagues in the pulsar group of Peking University for the helpful
discussion. This work is supported by NSFC (10778611, 10973002), the National Basic Research Program of China
(Grant 2009CB824800), and by LCWR (LHXZ200602).






\end{document}